\documentclass[manuscript,natbib=false]{acmart}

\AtBeginDocument{%
  }

\usepackage[english]{babel}
\usepackage{amsthm}
\usepackage{csquotes}

\usepackage{array,booktabs,tabularx,multirow, multicol,colortbl,hhline,makecell}

\usepackage{tablefootnote}

\usepackage{graphicx,xcolor,caption,placeins,wrapfig}

\usepackage[noabbrev,nameinlink,capitalize]{cleveref}

\RequirePackage[
  datamodel=acmdatamodel,
  style=numeric-comp,
  sorting=none,
  ]{biblatex}

\begin{document}

\title{Trustworthiness in Stochastic Systems: Towards Opening the Black Box}

\author{Jennifer Chien}
\email{jjchien@ucsd.edu}
\affiliation{%
  \institution{UC San Diego}
  \country{USA}
}

\author{David Danks}
\email{ddanks@ucsd.edu}
\affiliation{%
  \institution{UC San Diego}
  \country{USA}
}

\setcopyright{none}

\begin{abstract}
AI systems are increasingly tasked to complete responsibilities with decreasing oversight. This delegation requires users to accept certain risks, typically mitigated by perceived or actual alignment of values between humans and AI, leading to confidence that the system will act as intended. However, stochastic behavior by an AI system threatens to undermine alignment and potential trust. In this work, we take a philosophical perspective to the tension and potential conflict between stochasticity and trustworthiness. We demonstrate how stochasticity complicates traditional methods of establishing trust and evaluate two extant approaches to managing it: (1) eliminating user-facing stochasticity to create deterministic experiences, and (2) allowing users to independently control tolerances for stochasticity. We argue that both approaches are insufficient, as not all forms of stochasticity affect trustworthiness in the same way or to the same degree. Instead, we introduce a novel definition of stochasticity and propose latent value modeling for both AI systems and users to better assess alignment. This work lays a foundational step toward understanding how and when stochasticity impacts trustworthiness, enabling more precise trust calibration in complex AI systems, and underscoring the importance of sociotechnical analyses to effectively address these challenges.
\end{abstract}

\maketitle

\section{Introduction}
Trust is critical for successful, sustained deployment and use of complex technological systems~\cite{alhogail2018improving, nadella2023understanding, akinwunmi2015trust, benbasat2005trust, jensen2015trust}. Trust is even more critical for present-day AI systems, which are increasingly deployed in settings with decreasing human oversight or opportunities for feedback. In contrast with mere reliance, predictability, or stability, trust implies a willingness to make oneself vulnerable to the risk of undesired consequences, often by foregoing alternative or parallel means of completing a task~\cite{arif2022vulnerability}. Trust -- when justified -- requires that the trustor has appropriate expectations that the system will behave in the ``right'' ways from the trustor's perspective. These expectations imply a degree of alignment between the trustor's values and those embodied by the system. 

Trust and AI have been widely discussed from a range of different perspectives. Our analysis here is prompted by a specific challenge to trustworthiness salient in most modern AI systems: stochasticity. In this case, one might naturally wonder whether it is possible to have (warranted) trust in a system whose behavior can shift, even when it is repeatedly placed in the same context. More generally, stochasticity may raise a myriad of concerns, ranging from inappropriateness in high-stakes domains to reproducibility~\cite{chien2023algorithmic, hill2024machine, impagliazzo2022reproducibility, ahn2022reproducibility}. And if the trustor cannot determine that the AI system will behave in alignment with their values, then why should they make themselves vulnerable to it? 

Stochasticity risks a particularly dangerous kind of untrustworthiness in which a system that is normally trustworthy may (stochastically) act in ways that open up users to risk. If that stochasticity is undetected and unmitigated, then the risks can potentially be catastrophic. For example, if a self-driving car operates safely 99\% of the time with no indications of when it will not, then I might inappropriately trust the car to \emph{always} behave appropriately, thereby leading me to disregard safety precautions and mechanisms for recourse (e.g. override features)~\cite{Goldman_2024, Thadani_2023}.

We contend that the issue is even more complicated. Not all stochasticity is equally indicative of a system's untrustworthiness, as some forms of stochasticity may not matter to a user. For example, I might not care about the specific words selected for an output as long as the text has the ``intended'' overall meaning. In fact, stochasticity can sometimes be a positive factor depending on the user's values, such as in creative tasks and representations of diversity and uncertainty\cite{hauhio2024enhancing, vasconcelos2024generation, bradley2023quality, zhang2024generative, jain2023contextual}. We consider the complexity of what designates a ``stochastic'' system in more depth below, but note that values seem to be playing an important role in whether stochastic behavior is problematic for trust. This raises the question: How do I know when a system is in alignment with my values, and how do I approach trust in a system that may only sometimes act in alignment with my values?

In this work, we take a more theoretical or philosophical approach to analyze trustworthiness in stochastic AI systems. We begin by providing definitions for trust, trustworthiness, and stochasticity (\cref{Sec::Trust}), before illustrating how stochasticity complicates typical methods of establishing trustworthiness (\cref{Sec::Conventional Approaches to Trust}). We then consider two practical approaches to managing stochasticity: eliminating user-facing stochasticity to create deterministic experiences, or allowing users to independently control tolerances for stochasticity (\cref{Sec::Dealing with Stochasticity}). We argue that both approaches are ultimately insufficient, as not all ``behavioral noisiness'' equally affects trustworthiness. 

Instead, we suggest that we need a more nuanced understanding of stochasticity, and so we provide a novel definition of stochasticity tailored to its implications for trustworthiness (\cref{Sec::Value_Alignment}). We then develop a sociotechnical approach~\cite{lazar2023ai, selbst2019fairness} for latent value modeling (\cref{Sec::Opening the Black Box}) to determine the trustworthiness of a stochastic system. In our approach, we model latent values for both the system and users in parallel, allowing for a more precise assessment of value alignment in AI interactions. This work offers a foundational step toward understanding when and how stochasticity affects the trustworthiness of AI systems. By delineating the conditions under which stochasticity may or may not undermine trust, we contribute a reflexive tool and framework for enhancing trustworthiness determination and calibration in complex, autonomous AI systems.

\section{Defining Trust, Trustworthiness, and Stochasticity}
\label{Sec::Trust}

\subsection{Trust}
At a high level, we understand ``A trusts B'' to mean that (i) A [the trustor] makes themself vulnerable (ii) because of justified expectations that (iii) B [the trustee] will act in ways that support the relevant\footnote{That is, those values that are potentially impacted by the vulnerability.} values of A. We thus characterize trust in a normative way as a set of beliefs, decisions, and/or actions that underlie the willingness to depend on another party in situations with uncertainty or risk~\cite{tullberg2008trust, dasgupta2000trust, sekhon2014trustworthiness}. This understanding is consistent with many conceptions of `trust' found in philosophy, social psychology, retail, business, marketing, organization science, and more~\cite{rousseau1998,baier1986, kharouf2014building, sekhon2014trustworthiness, manzini2024should}. 

Trust differs from mere predictability, as trust is bidirectional while dependence based on predictability is only unidirectional~\cite{lyons2019trust, robbins2012institutional, zonca2021does}. If A merely relies on B (i.e., without any additional contracts or promises), then B does not have any obligation (ethical or social) to adjust their behavior to support A. In contrast, if A trusts B, then B should act as A would, if A had B's capabilities. That is, B's behavior should depend on (their understanding of) A's values, even as they may change with time or context. It is not sufficient for B to act in ways that accidentally happen to benefit A; rather, B must intentionally act to support A's relevant values.

This conception of trust is not a strict binary, but rather falls on a continuum, as the trustor might trust to a greater or lesser degree. Trust is also temporally sensitive, being built upon or eroded through interactions. We do not formalize a threshold for declaring ``A trusts B,'' as these are likely subjective or culturally specific~\cite{nishishiba2000concept, kwantes2021contextual, sofer2017your, kiyonari2006does}. Nonetheless, there are many situations in which A makes themself sufficiently vulnerable for the right reasons to warrant the description `trust'. We primarily focus here on cases in which the trustor is a person/user and the trustee is an AI system.

Trust ought not be uninformed nor reckless: warranted or justified trust is based on appropriate understanding of the potential negative consequences of the trust (e.g., risks relevant for the trustor) and perceived value alignment. A may make the decision to trust after receiving signals (i.e., information or observations) of the values that inform B's behaviors. These values indicate the manner in which B may make future decisions without either party having to specify the exact choices they intend to make. For instance, suppose A trusts B to draft a cover letter according to some mutually agreed upon values of authenticity, accuracy, and professionalism. A may not have to specify the exact format, word choice, or achievements required for a satisfactory result. Instead, B is trusted to execute these actions and make decisions as appropriate. Any precautions to protect against the negative consequences of a dissatisfactory result (and therefore inappropriate trust), such as additional rounds of reviews or editing, may be in proportion to the consequences of a poor letter.

\subsection{Trust in Technological Systems}
\label{Subsec::Trust in Technological Systems}
Technological systems raise distinctive issues for trust, as the discussion in the previous subsection suggests that the trustee might require complex mental or emotional states that AI (in particular) might lack~\cite{ryan2020ai}. More specifically, for a technological system trustee (B), what does it mean for that system to understand and be responsive to A's values? Since technological systems are inanimate and therefore unchanged by the act of trust, one natural move is to expand the scope of the system to include broader sociotechnical elements. That is, trust in the system would be partially grounded in trust of the designers, developers, regulators, and others (collectively) to ensure that the system would support my values. ``A trusts B'' thus means, on this analysis, that I trust that the relevant sociotechnical system (as a complex entity) will act---either individually or as a collective---to support alignment with, and continue to advance, my (relevant) values.

While this analysis works from a metaphysical or conceptual perspective, it faces enormous epistemic burdens. On the user side, I may not be consciously aware of all of my relevant values, may be unable to communicate them, or may change them with time or context~\cite{pugh1976human, curry2022moral, jin2012stability, bardi2009structure, buenger1996competing, schwartz1992universals}. Furthermore, each of the various actors in the sociotechnical system may similarly not know, be able, or willing to disclose the values embodied by their contribution to the system and its outputs. Nor may it be a trivial task for them to isolate which stakeholder is responsible for addressing and supporting a given value. Moreover, this approach would require that the sociotechnical system be able to detect and take relevant measures to respond to my values. This would likely require extensive (and invasive) amounts of data over a (potentially prohibitively) long period of time. Trust involving a sociotechnical system is therefore unlikely to be established through explicit elicitation and alignment of values.  

Alternatively, we can consider trust relative to a user and a particular technological system in a particular context. By narrowing the scope, we can thereby hold fixed the values supported by a technological system. We can then ask which users (or subset of users) have values appropriate for that particular system in that context. Rather than expecting the sociotechnical system as a whole to adapt to an arbitrary user's values across all contexts, we can potentially determine the values advanced by the system, and then adjust the corresponding subset of users who can justifiably trust it to support their values in that context.

\subsection{Trustworthiness}
While trust is the occurrent relation between two entities, trustworthiness is a property of the (potential) trustee, with respect to individuals (potential trustors) with specific values in specific contexts. In particular, B is trustworthy for A when there is appropriate value alignment such that A should trust B~\cite{sekhon2014trustworthiness}. Many different individuals might ``fit'' the A role, as long as their values would be appropriately supported by B. Importantly, trustworthiness is only when it is reasonable or defensible for A to trust B; it does not imply that A actually trusts B. More generally, there is an important distinction between when one \emph{does} trust as opposed to when one \emph{should} trust. In this paper, we focus on the normative notions of the latter: when one ought to trust and when a system ought to be deemed trustworthy.

\subsection{Stochasticity}
The term \textit{stochastic} generally refers to something that is noisy, random, unpredictable, or otherwise indeterministic. Here, we define a stochastic system in terms of its (technical) components: a system that incorporates one or more stochastic or probabilistic processes~\cite{kulkarni2016modeling}. Importantly, this definition of stochasticity is independent of whether a user can detect any variance in the system's behavior. It includes systems with probabilistic processes at any stage of development, including training, model selection, and output generation. We include such a general definition because of the potential generality of a user's values (i.e., we place no restrictions on how particular values may be specified). A user can have values as high-level as general characteristics for any output they receive, or as low-level as a specific numeric prediction given a particular input. 

For example, consider a user prompting a generative image model with the input, ``Show me a grilled cheese.'' We expect that there could be significant variation across some dimensions of the output image (e.g. bread texture, kind and proportion of cheese, direction of cut, etc), even if the system reliably outputs an image of a grilled cheese sandwich. This generative image model is thus stochastic. Notice that stochasticity is a property solely of the technical system; features of the user are not relevant to whether the system is stochastic. By our definition, the stochasticity could arise in many ways. A system that selects a random seed to determine the starting point for surveying training images would qualify as stochastic. Similarly, a system in which multiple probabilistic processes (e.g., initialized weights and dropout in neural networks) are combined to produce a final outcome also qualifies as stochastic. Finally, stochasticity could arise because the system generates outputs by sampling from a distribution of possible outcomes, where one cannot guarantee a particular output or uniformity in the outputs (for a fixed input). 

\section{Stochasticity and Conventional Approaches to Trust}
\label{Sec::Conventional Approaches to Trust}
\subsection{Stochasticity may Undermine Trust}
Trust depends not just on the system supporting a trustor's values on one occasion, but rather on it doing so consistently. Stochasticity can undermine trust by introducing variance in the system's behavior, which may lead to inconsistent support for values relevant to the trustor. This intermittent misalignment makes it difficult for the trustor to form justified, stable expectations about the system, relative to the core risks.

The ``sometimes'' nature of this misalignment exacerbates efforts to determine trustworthiness precisely because it creates uncertainty. For example, consider an AI system that assists with medical diagnoses. Suppose the system accurately aligns with a patient's need for safety and precision in most cases, but occasionally suggests a diagnosis inconsistent with medical best practices due to stochasticity. In this case, the potential trustor (patient or healthcare provider) will, if they learn about the stochasticity, be left questioning whether the system's outputs can be trusted in any given situation. Even if the overall frequency of misalignment is low, the fact that such events are unpredictable means the trustor must remain vigilant, undermining their confidence in the system and the practical value of trust~\cite{passi2022overreliance, kim2024m, zhang2024you, sadeghi2024explaining, vasconcelos2023explanations}. Intermittent misalignments can be especially insidious if they provide the illusion of trustworthiness while still failing to consistently uphold the trustor's values. This pattern disrupts the predictability and dependability required for trust, leaving the trustor uncertain about whether the system will support their values at any given moment. As a result, the trustor may either overestimate the system's trustworthiness (based on its generally good performance) or become overly cautious and withdraw trust entirely (because of its occasional failures).

This sporadic variance also complicates trust-building through observation or interaction. A trustor might initially experience many positive interactions, leading them to infer that the system aligns with their values. However, when a single instance of misalignment is detected, this may erode trust and may further create doubt about the trustworthiness of all prior and future outputs. This is particularly troubling in high-stakes contexts, where even infrequent deviations from expected behavior can have significant consequences. Below we list some of the ways in which stochasticity further complicates canonical approaches to building trust if the technical system is treated as a black box, leaving trustors to infer system values based on outputs or external signals alone. 

\subsection{Conventional (Black Box) Approaches to Trust}
For clarity and ease of reference, we introduce a running example to demonstrate and concretize the potential problems that may arise. Consider an interaction with a generative AI system (e.g., a large language model (LLM)). Generative AIs are increasingly used in diverse, high-stakes applications where trust and value alignment are crucial~\cite{ramdurai2023impact}. Even if a human is ``in'' or ``on'' the loop, they might use the generated outputs uncritically, thereby inducing the kinds of vulnerability that trust is supposed to address. At the same time, due to their probabilistic nature and inherent complexity, generative AIs often exhibit stochastic behaviors that can either uphold or undermine user trust, depending on how well their outputs reflect user expectations and values. We will focus on the specific case where a user prompts a generative image model with the input, ``Show me a grilled cheese.'' This model contains stochastic processes at various points throughout development and deployment: partitioning and sampling of training data, expert availability at query time, selecting outputs (e.g., those equivalent or similar in the eyes of the model), and more. We thus consider the question: When could one justifiably trust this generative AI system (in response to this prompt)? 

\subsubsection{Observation \& Interaction}
One of the first and most intuitive ways one might go about establishing trust is through observation and interaction, as these can provide one with information about the map between the system's actions and the values driving its behavior. Preliminary interactions with a system may reveal high-level capabilities, such as ability to understand and comply with requests, as well as impute and resolve appropriate ambiguities. These interactions alone might be sufficient to know whether to trust the system. For example, if the system outputs an image depicting two grilled slices of cheese sandwiching a kitten, then I can conclude that the values that inform this model's outputs (e.g., what constitutes a sandwich) do not align with my own. 

The success of this approach, however, rests on some restrictive assumptions. First, it assumes that the system's design reflects a consistent set of values and that its actions reliably signal these values. In deterministic systems, this is already a high bar, as design decisions may not be driven by a consistent value set nor consistently represent developer-intended values~\cite{bauer2009designing, molina1997insights, macdonald2009preference, arvan2024interpretability}. For stochastic systems, values and risks of stochasticity might not be considered by the developers, or they might even view stochasticity as a feature rather than a bug. 

More importantly, actions are typically only noisy signals of, and may not uniquely be determined by, the underlying values. Stochasticity further complicates this relationship, introducing variability in behavior for the same input, additionally obscuring the mapping between actions and values. For individual users, discerning underlying system values can be practically infeasible (in some reasonable amount of time and effort), as the variability in outputs makes this inference difficult. In other words, it may be unclear the extent to which an individual user is able to draw conclusions about general system capabilities and values based on their individual experiences alone. An ``untrustworthy'' system's behavior may vary only minutely over extended periods of time, say if the model only outputs something unusual once in every 10,000 outputs. Establishing trust is already a difficult challenge in deterministic systems, but it is therefore rendered significantly harder for stochastic systems given the increased interaction requirements to make capability or value judgments. 

Alternatively, one may choose to rely on an auditor or quality assurance team (rather than individual users) to determine system values from observations of behavior. For example, data scientists might employ probabilistic methods like Bayesian inference to estimate the likelihood of value alignment based on behavioral distributions (i.e., most probable value set given some set of behaviors and interactions). This posterior distribution could also be used to calculate the probability that a system's behavior will fail to support the values of a user by taking the expectation over the distribution. More generally, auditors, regulators, or others may be able to take longer-term or wider-scope approach that offsets many of the concerns about limited experiences for a single user.

However, values can manifest in complex, context-sensitive ways; a system may support a high-priority value in one setting that may be irrelevant in another. Just because a system seems to exhibit value alignment for sandwich images, does not mean it will do so in producing images of people~\cite{Vynck_Tiku_2024}. As a result, the behavioral distributions (conditional on values) are likely to be extremely complex, if not impossible to specify. The addition of stochasticity will likely only increase the number of observations and interactions required for reliable inference (from only observation and interaction) of the values that the system supports, posing computational feasibility challenges.

\subsubsection{Social \& Technical Roles}
Systems may also be trusted based on their social or technical roles, similarly to how people are sometimes trusted in virtue of their professional (or other) role. As a non-AI example, a basis for trust in one's doctor, even on the first visit, is a reasonable expectation that they will conform to their social role by focusing on the patient's health. For sociotechnical systems, this role-based trust may be grounded in expectations such as certifications, governance, or professional standards. This approach to building trust is thus limited to cases in which such structures exist; for example, a generative image model would only have an appropriate role in some domain application, such as for medical health diagrams or food advertising. If there are no relevant laws, rules, or standards (whether legal or social) in that domain,\footnote{The extent to which technology should or should not inherit all regulation from the application domain it is applied to is an open question (see e.g., credit scoring and third-party data usage in insurance~\cite{kiviat2019moral, andrews2023algorithmic}).} then there will likely not be an appropriate role that can ground the trustworthiness. For many AI systems, the requisite social or technical roles do not exist~\cite{al2023review}. 

In addition, most technical and social roles (at least, for people) involve an expectation of deterministic behavior (e.g., give the best treatment for the patient's condition), even if there may be variance across what the best treatment is. Stochastic systems have the potential to exhibit variability in their behavior; for example, they might assign a different treatment to the same patient depending on when the diagnosis is requested, or assign different chances of success for patients who are viewed as biomedically similar. Either behavior would make the system less likely to appropriately fill such roles~\cite{quinn2021trust}. Stochastic systems thus present a significant challenge to this form of trust-building. Variability by stochastic systems can lead to inconsistent fulfillment of role-based expectations, undermining their ability to fit established roles. For our running example of an image generator, even if it were specified to generate content for food advertising, there may be stochasticity that manifests as sometimes valuing creativity (e.g., abstract food concepts and representations such as in molecular gastronomy or synesthesia) and otherwise valuing naturally occurring depictions of food (e.g., how the food would actually be sold to the consumer).

Furthermore, stochasticity exacerbates governance challenges. Variability can hinder clear assessment and regulation, leading to problematic feedback loops. Widespread system usage might reinforce trust in regulatory approval, while regulators might continue approval due to apparent public acceptance. This circular reliance erodes the trustworthiness of the system and its governance structures, particularly when stochasticity produces inconsistent outcomes. Suppose an image generator becomes the standard system for stock food photos, to the point that people no longer use actual stock photos of the product. The widespread trust that restaurants and grocery stores have entrusted in using these models and their outputs, along with corporate lobbying, may reduce public and regulator knowledge and scrutiny towards regulating these products. Examples of this have been seen in regulation of the internet and social media such as Facebook (Meta) and Twitter (X)~\cite{siapera2022platform, davidson2024argumentation, popiel2018tech, kourabas2023big}. 

\subsubsection{Transitivity}
Transitive trust occurs when individuals rely on the trust or experiences of others, especially when they believe others share similar values. For example, if my friend trusts a system to handle sensitive information or produce images for a work presentation, then I might also be inclined to trust it. More generally, values shared by most people (e.g., valuing personal health) may mean that social proof--doing or believing whatever is done or believed by many others--can be an effective mechanism to build warranted trust~\cite{cialdini2001harnessing}. However, stochasticity can undermine this process by increasing uncertainty: I do not know whether the system will behave similarly for me (even if I am in the same context as my friend), and so their trust is not necessarily informative for me. More generally, a stochastic system's variability may make it harder to justify trust based on others' experiences. 

We have argued in this section that, in all three approaches, stochasticity complicates the alignment of probabilistic processes with users' values, disrupts traditional methods for building trust, and highlights the difficulty of discerning trustworthiness. Next, we consider two natural responses to the tension, where the proper response depends partly on who bears the responsibility (e.g. developers, users). Although these are both being pursued in current AI systems, we identify challenges to both approaches. We then turn in Section~\ref{Sec::Value_Alignment} to our proposed innovative framework that accounts for the nuanced impact of stochasticity on trust-building mechanisms, including latent value modeling and sociotechnical analyses to better assess and manage stochastic systems.

\section{Responses to the Tension}
\label{Sec::Dealing with Stochasticity}

\subsection{Eliminating Stochasticity}
One way to reduce the potential tension between stochasticity and trustworthiness is to eliminate variability at the point where users encounter it; for example, we might require all outputs of grilled cheese images to be the same for some query. This effectively ensures that consumers consistently receive the same response for a given input. By doing so, stochastic systems are effectively rendered deterministic from the user's perspective, allowing conventional approaches to trust-building to be applied. This can simplify users' expectations of the system, as they know that similar inputs will consistently yield similar outputs, reducing the uncertainty that could undermine trust.

However, while this method may increase perceived trustworthiness, it also risks removing potentially beneficial or valuable aspects of stochasticity. For instance, variability in responses can enable users to explore diverse perspectives or options without having to alter their input, providing flexibility and a broader set of possibilities~\cite{lee2024one, kim2024m}. This variability can be particularly useful in domains like creative problem-solving, language generation, and recommendation systems, where variability in outputs can enhance user satisfaction and lead to more diversified results~\cite{kim2021customer, ashkinaze2024ai, zhu2023role}. By eliminating \emph{all} outward-facing stochastic elements, systems may inadvertently constrain user experience, offering only a narrow set of responses that limit exploration and creativity~\cite{liu2024chatgpt}.

Deterministic behavior can also inadvertently convey an illusion of comprehensiveness or certainty, which may be misleading, particularly in complex or open-ended tasks~\cite{kidd2023ai, kosch2023placebo, dreyfuss2024human, chien2024beyond, sun2024ai, duede2024paradox}. Users might assume that the fixed output represents the ``best'' or ``correct'' answer when, in fact, the output could be one of many valid responses. This can hinder users' ability to critically evaluate outputs, potentially leading them to overly rely on the system without considering alternative viewpoints or verifying its outputs.

Thus, while rendering a stochastic system deterministic may appear to increase trustworthiness, it is not always desirable or effective. The challenge is to balance the need for consistent, predictable behavior with the benefits of output diversity that users may find valuable.  

\subsection{Representing Stochasticity to Users}
Instead of eliminating stochasticity outright, an alternative approach might be to directly communicate the variability to a user, allowing them to assess whether a system is likely to operate in a trustworthy manner. That is, system designers could embrace stochasticity as an inherent feature while they give users control over the degree of output variability~\cite{kay2016ish, wang2024enhancing}. This control might be implemented as one or more ``control dials'' that directly allow a user to specify the temperature, entropy, or variance embedded in their outputs. This approach places the responsibility on users to fine-tune or predefine their preferred level of randomness based on their individual needs, allowing for a customized balance between predictability and flexibility. Ideally, this control could empower users to determine how much variability they want in responses: they might desire consistent, deterministic answers or instead be open to a range of responses that reflect the underlying randomness in the model.

However, in its simplest form, treating all forms of stochasticity uniformly may oversimplify the complexity of stochastic behavior in AI systems. If users are presented with only a single control mechanism to adjust the overall variability, then the system would treat every aspect of output generation as equally variable. For instance, this approach might not differentiate between stochasticity in interpreting user intent and in selecting output components, treating both as part of a single ``variability'' knob. In practice, these are distinct elements: input interpretation affects the understanding of user queries (e.g., what qualifies as a sandwich), while variability in output selection affects the specific content of the response (e.g., the size of the sandwich). Combining all of these potential axes of variation under a single control is likely to be both impractical and unintuitive for users~\cite{hullman2019authors}, as they may want more control over one form of stochasticity than the other.

At the most complex end of the spectrum, users could fine-tune multiple layers of stochasticity at various points in the input-output pipeline, though at the cost of significant, and perhaps overwhelming, complexity. Users might need to adjust settings for stochasticity during input interpretation, content generation, and output selection, each with its own specific control. This could lead to a burdensome level of customization~\cite{stern2014improving}, requiring extensive trial and error and a deep understanding of each layer's role in shaping the output, particularly in domain- or task-specific occurences~\cite{bodrunova2018impact, findlater2004comparison}. Moreover, downstream interactions between these layers of stochasticity may produce unpredictable results despite the user's best efforts.

This level of involvement is likely undesirable and impractical, especially for users who expect straightforward, reliable, and intuitive interactions. Requiring users to make adjustments across multiple layers of the system may deter engagement and complicate the user experience. Users may be overwhelmed by the complexity of options and frustrated by the need to experiment to achieve the right balance of stochasticity. In some cases, these challenges could diminish users' trust in the system, as they may perceive the need for extensive tuning as a sign that the system is inherently unstable or unpredictable. Thus, while giving users control over stochasticity may seem beneficial in theory, its implementation may introduce new usability and trust challenges that ultimately undermine the desired benefits.

\section{Proposal: Value Alignment for Trustworthiness}
\label{Sec::Value_Alignment}
Given the limitations of these two approaches for managing stochasticity, we propose value alignment as a framework for addressing this challenge: if the system's and user's relevant values are aligned for a given task, then the system is trustworthy, regardless of stochasticity. In the following section, we propose a new definition for stochasticity that focuses specifically on its potential threats to value alignment, before delving into two current approaches. We outline the strengths and weaknesses of both of these approaches to motivate our proposed alternative approach: a sociotechnical approach to latent value modeling (\cref{Sec::Opening the Black Box}). That being said, all three approaches imply the same key takeaway: value alignment is at the core of the trustworthiness of an AI system. 
\subsection{Stochasticity (Refined)}
We start by refining our previous definition of stochasticity to capture the idea that only some kinds of stochasticity ``matter'' (particularly when thinking about trust). If a system is stochastic in a way that does not matter for the user's (relevant) values, then this stochasticity does not actually present a challenge to the system's trustworthiness (in that context, for those values). For example, if I only care about my image generator producing a picture of a grilled cheese sandwich, then stochasticity about the type of bread will (likely) not impact trustworthiness. Our previous definition identified stochasticity broadly as any system incorporating probabilistic processes, regardless of whether a user could perceive the variance. By contrast, our refined definition focuses on whether there is meaningful variation relative to one's values. Specifically, we define a \textbf{\textit{stochastic system}} as one that produces outputs (given a fixed input) that exhibit variability (i) at or above the user-specific \textbf{level of relevant description} (as determined by their values and input), or (ii) beyond their \textbf{knowledge}. This shift prioritizes understanding stochasticity's role in undermining trust by focusing on its alignment (or misalignment) with user values.

Variation \textit{at} a user-specific level of relevant description pertains to outputs conflicting with those directly specified by the input or the relevant user values.\footnote{More specifically, relevant values determine the ``language'' in which an event or situation should be described, and thereby determine a corresponding level of description.} For instance, if a user asks for a grilled cheese sandwich and receives an image of an uncooked cold sandwich, the system has failed to uphold alignment \emph{at} the level of description explicitly provided by the user. If a sandwich image generation system produces two pieces of grilled cheese on either side of a kitten, then we have variation \textit{above} a level of description since the very concept of a sandwich varies.

Furthermore, variation outside a user's level of knowledge may also indicate stochasticity. Consider a sandwich-making robot that ceases to function at high temperatures, a limitation unknown to the user. While this behavior may be deterministic from the system's perspective (when temperature is an explicit input), it appears stochastic to the user because their knowledge does not account for this factor. \footnote{For reasons of space, we do not explore questions about what users actually know versus what they ought to know. Similarly, we also leave aside questions of which values a system should impute when values are not explicitly specified.} Consequently, the perception of a system as stochastic depends not only on its technical processes but also on its interaction with user-specific knowledge, user-relative levels of description, and expectations.

With this refined definition, we transition from considering the presence of probabilistic processes to focusing on their practical relevance to value alignment and trustworthiness. By centering the definition on a user's level of relevant description and knowledge, we gain a clearer framework for identifying when stochasticity poses a challenge to trustworthiness. With this understanding in place, we now consider various approaches to achieving value alignment, exploring how systems can ensure their behavior consistently reflects and upholds a user's relevant values, even amidst technically stochastic processes.

\subsection{Direct Implementation}
One approach requires identification or elicitation of user values prior to design and development, and then directly implementing those values into the technological system. This method faces a myriad of challenges that can be categorized into human and technical challenges.

Human values are highly complex, context-dependent, subjective, dynamic, and often contradictory~\cite{curry2022moral, jin2012stability, bardi2009structure, buenger1996competing, schwartz1992universals}. Users may not be consciously aware of all their values, making it difficult to articulate them clearly~\cite{pugh1976human}. Additionally, there is no universal agreement on how to prioritize or interpret values such as fairness or justice, which can vary significantly across cultures, communities, and individuals~\cite{jacobs2021measurement, dator2006fairness}. This variability can lead to potential misalignment with ethical or moral outcomes when implementing values without nuanced understanding.

Users may also need to make difficult trade-offs between competing values, such as balancing privacy with transparency or safety with freedom. These trade-offs can become overwhelming and complex, especially when applied across different applications and scenarios. Systems perceived as enforcing values in a rigid or prescriptive manner may not align with user expectations and may fail to accommodate diverse perspectives. For example, a system that prioritizes safety at all costs may require all users to fully disambiguate all aspects of a query: size, ingredients, sandwich orientation, kind of cheese, inclusion of crust, direction of grill marks and cutting, color tones in background, and so much more. All of this specificity is an attempt to minimize potential imputation of elements that increase likelihood of undesired cognitive effects. However, without consideration of contextual nuances, this can lead to overly restrictive behavior that limits user autonomy or infringes on freedom (e.g., never allowing a user to generate a sandwich with a happy face to mitigate likelihood of anthropomorphism).

From a technical standpoint, implementing human values directly into a system is difficult due to the complex, evolving nature of values. As values shift over time, systems would require ongoing re-implementation and updates to stay aligned. This places a significant burden on developers to continuously adapt the system to reflect these changes accurately. In addition, the need for systems to consider nuanced interpretations of values also poses a challenge. Designing a system capable of handling such complexity while allowing for context-dependent applications is a substantial technical hurdle. Ensuring that users can input their values and specify trade-offs in a manageable way adds to the difficulty. Without careful design, the system may become impractical and unintuitive for users, complicating its effectiveness and user satisfaction.

\subsection{Behavioral (RLHF)}
Behavioral approaches seek to bypass direct value elicitation by using limited forms of feedback, one of the most prominent methods being Reinforcement Learning from Human Feedback (RLHF). RLHF involves using human feedback to train models, allowing them to estimate human preferences and adapt the system behavior accordingly~\cite{bai2022training}. Through this process, a reward model is adjusted using human responses to certain outputs, guiding the system to align its responses more closely with human expectations and preferences without explicitly coding values into the system.

However, RLHF and similar approaches are inherently constrained by their behaviorist methodology. They focus on refining the system's outputs to match user feedback but are limited to what users can directly specify or articulate. This limitation means that RLHF and similar approaches may struggle to capture deeper, implicit, or more complex value structures that users may not be consciously aware of, or even be able to express. The reliance on observable feedback restricts the system's learning to the preferences that users can tangibly demonstrate, leaving broader, contextually nuanced, or less overt aspects of value alignment unaddressed~\cite{chaudhari2024rlhf, khamassi2024strong, casper2023open}.

Moreover, the feedback-driven nature of these approaches introduces potential biases and inconsistencies. Users' feedback may reflect momentary preferences or immediate reactions rather than holistic or principled value systems. As a result, the system may learn to optimize behavior based on surface-level feedback, which might not align with long-term ethical standards or underlying moral beliefs. This behaviorist approach can lead to a system that performs well in narrow, specific situations but fails to generalize its behavior in a way that reflects the more comprehensive, multifaceted nature of human values.

\section{Opening the Black Box: A Sociotechnical Approach to Latent Value Modeling for Trustworthiness in Stochastic Systems}
\label{Sec::Opening the Black Box}
In light of the challenges associated with black-box methods and the aforementioned approaches to value alignment, we propose an alternative strategy: opening the proverbial ``black box,'' adopting a sociotechnical perspective to identify, isolate, and evaluate the values embodied by a system's outputs. Specifically, we propose a latent value-modeling\footnote{Note that we are modeling human values that happen to be latent, rather than modeling the values of latent variables.} framework to capture both user and system values to determine alignment in stochastic systems. In particular, while users may not be concerned with the granular specifics of which values influence each stage of the system's pipeline, understanding these values in finality is essential to evaluating whether the system's overall embodied values align with user trustworthiness expectations. In this section, we apply our latent value-modeling framework to LLMs to demonstrate how this approach can illuminate the underlying values influencing outputs, thereby enabling more transparent and trustworthy interactions.

\subsection{Latent Value Modeling}
Our goal is to determine whether systems exhibit stochastic behavior that undermines trustworthiness for specific users. To achieve this, we propose latent value modeling---a framework that seeks to explicitly represent and disentangle the values underpinning decisions across the AI pipeline and throughout interactions. We argue that a decompositional approach is essential to determine trustworthiness, as the complexity of modern AI systems makes it infeasible for users to infer alignment solely from input-output mappings. More precisely, we need to identify and quantify how different stakeholders' values contribute to the system's overall embodied values and, ultimately, to the final outputs. By directly modeling the contributions of various stakeholders and technical processes, we enable users to better assess whether a system's stochastic variability impacts their values. At the same time, this framework empowers developers and model-owners to self-audit how values are embedded in their systems, providing a pathway for transparent accountability and reflexivity.

Consider the example of a user interacting with an LLM. The LLM is almost certainly stochastic in a technical sense, generating varied outputs for the same fixed input~\cite{ouyang2023llm, lee2024one}. However, trustworthiness hinges on whether the system exhibits relevant stochasticity: whether that variation in outputs results in inconsistent alignment with the user's motivating values. The latent value modeling approach requires that we explicitly ``open the black box'' to reveal and model the latent values embodied in the system because of choices across the development pipeline.

\begin{figure}[htp]
    \centering
    \includegraphics[width=4in]{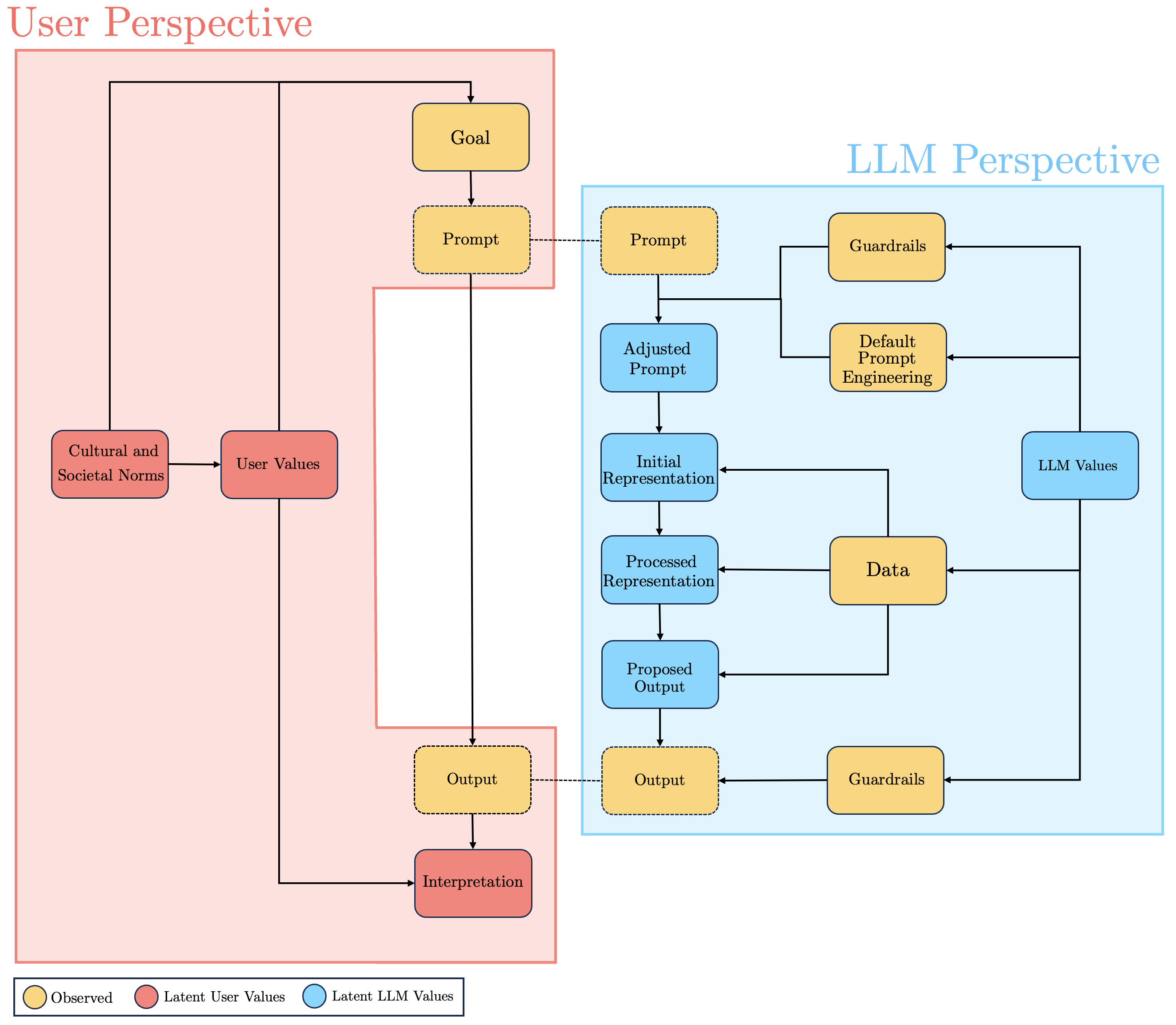}
    \caption{Causal diagrams from User and LLM Perspectives. Yellow nodes are observed elements, red are user-determined latent states and blue are LLM-determined latent states. We separately depict the user perspective (left) and the LLM perspective (right).}
    \label{Fig::Causal Diagram 1}
\end{figure}

Our proposed framework is illustrated in ~\cref{Fig::Causal Diagram 1}. In this model, the yellow nodes represent observed elements, such as a prompt or output, from the given perspective. The red and blue nodes correspond to nodes with latent states that we propose to model. The overarching goal is to better understand how values contribute to each half: for users from goal to interpretation and for LLMs from prompt to output. By understand the latent values contributing to both perspectives, we can then evaluate whether the values are sufficiently in alignment to designate trustworthiness. 

On the user side (red box), user values shape the interaction with the LLM, influencing their goal, prompt, and interpretation of the output. These values ultimately drive the user's expectations of trustworthiness. If users could directly discern the LLM's latent values -- such as those embedded in the system's training, guardrails, and intermediate processes -- they could more confidently assess whether the system aligns with their values. However, as discussed in ~\cref{Sec::Conventional Approaches to Trust}, users may struggle to infer the system values if there is stochasticity that occurs at or above their level of description, complicating this assessment and possibly eroding trust.

Complementary to this perspective, we introduce the LLM Perspective (blue box), which decomposes the LLM's embodied values into distinct components. This includes contributions from developer- or platform-defined guardrails (top and bottom instances), default prompt engineering, and data.\footnote{This causal diagram highlights how guardrails and default prompt engineering are currently observable causal factors -- often instantiated as rules (e.g., do not output toxic text or prepend instructions on a prompt). One could instead envision diagrams for future models in which the guardrails, default prompt engineering, and data nodes are also blue, signifying that they contain latent states that contribute values to downstream nodes.} These nodes reflect the points at which stakeholder values and technical decisions shape the LLM's behavior. By systematically analyzing how values influence intermediaries (e.g., prompt, adjusted prompt, initial representation, processed representation), we can identify where and how variability in outputs emerges, offering a reflexive tool and opportunity for rigorous interrogation and refinement of contributed values. While some of these steps (e.g., initial representation) may not directly correspond to explicit technical processes, they serve as conceptual points for isolating value influences within the causal pipeline.

\begin{figure}[htp]
    \centering
    \includegraphics[width=4in]{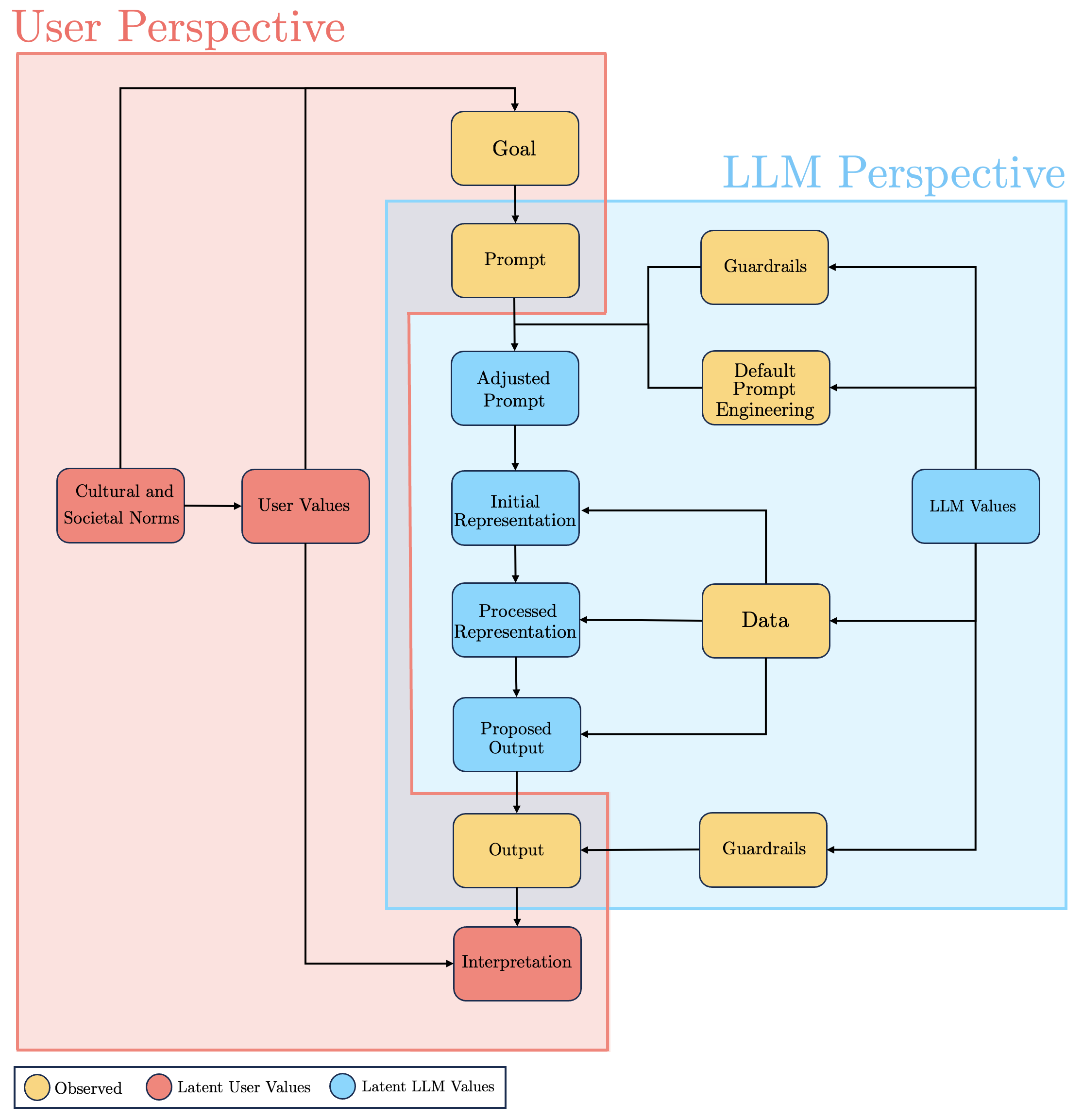}
    \caption{Causal diagram of User and LLM Perspectives Combined. Yellow nodes are observed elements, red are user-determined latent states and blue are LLM-determined latent states. We depict the user perspective and the LLM perspective overlapping in the central causal chain of the generative AI from prompt to output.}
    \label{Fig::Causal Diagram 2}
\end{figure}

Though the user and LLM perspective are depicted as separate views in ~\cref{Fig::Causal Diagram 1}, they actually represent the same causal system, unified by a central causal chain from prompt to output. We visualize this in~\cref{Fig::Causal Diagram 2}. The distinction in these two perspectives demonstrates the extent to which stakeholders may have different views of the observability of causal factors in the system: from the user perspective, their goal of interacting with an LLM is likely very clear, but it is not observable from the LLM perspective. This underscores the need for sociotechnical approaches that reflect the interaction between stakeholders on each side of the system.  

This causal model is intentionally simplified, omitting multiple connections (e.g., between cultural/social norms and guardrails). However, it suffices to demonstrate the critical importance of decomposing the causal chain from user values to system outputs. By making the contributions of each node explicit, this approach enables users and developers alike to identify whether stochasticity disrupts alignment with specific values. Crucially, it shifts the focus from treating the system as a ``black box'' to directly modeling the latent (moral, ethical, social) values that govern its outputs. Our framework underscores that not all stochasticity is inherently problematic for trust. The system's stochastic nature becomes a barrier to trust only when variability negatively impacts user values. By directly modeling these values, we can move toward more precise and actionable trustworthiness assessments in stochastic AI systems.

\subsection{High-Level Approaches for Inferring (Model) Values}
\label{Sec::Determine Values}
This causal modeling approach requires some way of determining the stakeholder values, and thereby understanding how choices (informed by values) impact the outputs of the system. We consider two different approaches to this challenge. First, an inference-based approach involves estimating the underlying values for each stakeholder. For example, we can treat the red and blue nodes as (mostly) unobserved variables, and then use any number of latent variable estimation procedures to determine the most probable values for each node~\cite{moon1996expectation, brooks1998markov}. These estimates can be used to predict similarity and appropriateness of outputs for different topics, domains, and cultural contexts. A fully-specified model (or set of models) can then be used to determine domains for which the model is trustworthy for users with specific value profiles. Of course, this approach requires a significant amount of data, and the space of latent variables must be compatible with (i.e., vaguely resemble) human values. For both of these reasons, this approach may not work, or may make interpretation difficult~\cite{arvan2024interpretability}. 

Alternatively, we can leverage control over specific technical elements of the system. By systematically modifying elements such as the scope of training data, the learning algorithm, or model complexity while holding others constant, we may be able to compute simplified, domain-specific value sets. This could enable us to establish trustworthiness for specific models, domains, and value sets. Such an approach may also take advantage of the relative uniformity in training datasets and underlying architectures across many large language models (LLMs){\cite{naveed2023comprehensive}. However, the influence of other nodes, particularly guardrails, on the composition of value sets introduces complexities that make direct comparisons across models challenging. Additionally, while this type of small-scale controlled testing could yield valuable insights, it is unlikely to generalize easily to the vast array of topics, domains, and deployment settings associated with present-day LLMs and, presumably future agentic AI systems.

\section{Conclusion}
In this work, we adopt a theoretical/philosophical perspective to explore the role of stochasticity in assessing the trustworthiness of AI systems. We argue that conventional approaches to trust fall short and that existing technical definitions of stochasticity are too broad to precisely identify its threat to trustworthiness. To address this, we outline three potential approaches to managing stochasticity: eliminating it, representing it to users, and aligning it with user values. For value alignment, we further distinguish between two current strategies: direct implementation and behavioral alignment via techniques such as RLHF. However, the limitations of these methods underscore the need for a more nuanced sociotechnical approach. To this end, we propose latent value modeling, a framework for identifying and interpreting the values embedded throughout the AI pipeline. This approach represents a significant step toward developing interpretable and trustworthy stochastic AI systems, while also raising critical challenges in value inference. By addressing these challenges, our work lays the foundation for a deeper understanding of how to align AI systems with human values in complex, dynamic contexts. 

\newpage
\printbibliography

\newpage
\appendix

\section{Stochasticity and Trustworthiness in Interpersonal Relationships}
\label{App::Stochasticity and Interpersonal Trust}
Stochasticity in interpersonal relationships is far less likely to undermine trust than in technical systems for several reasons. First, human values tend to be relatively stable over time and change gradually rather than randomly\cite{milfont2016values, vecchione2016stability, dobewall2016rank, manfredo2016implications}. Sudden, unpredictable shifts in a person's values are exceedingly rare and often occur in response to significant external events~\cite{daniel2022changes, gouveia2015patterns, bardi2009structure, lonnqvist2011personal, bojanowska2021changes, daniel2013brief, sortheix2019changes, verkasalo2006values}. This stability provides a consistent foundation for trust, as individuals generally expect others to act in accordance with their established principles and behaviors. In contrast, technical systems, particularly those with stochastic elements like LLMs, can exhibit output variability due to  probabilistic processes, creating unpredictability that undermines user trust.

Humans often signal their intentions, preferences, and even changes in values through explicit communication or implicit cues, allowing others to anticipate or adapt to shifts in behavior~\cite{lindenberg2000takes}. When someone's priorities evolve, these signals enable recalibration of expectations, thereby preserving trust. Interpersonal relationships also operate within shared cultural, social, and emotional contexts that provide interpretive frameworks for understanding behavior. Even when someone behaves unpredictably, shared norms and values can help contextualize their actions as intentional or meaningful rather than random. Technical systems, however, are often lauded for their lack of context restrictions, meaning that it often lacks the capacity for signaling such context-specific dependencies~\cite{goertzel2014artificial}. This means that when a system produces a response that deviates from prior behavior when deployed in new contexts, it does so without warning or explanation, leaving users unable to predict or understand the variation. This absence of transparency has the potential to make technical stochasticity feel riskier and more destabilizing to trust.

Accountability further differentiates human relationships from technical systems. Humans are guided by social norms, reputational consequences, and emotional bonds that discourage inconsistent behavior and promote trustworthiness. In contrast, technical systems are not inherently accountable, as their behavior is governed by their design and programming rather than relational dynamics. Their context-freeness only magnifies the lack of clear accountability to accurately determine the risk of stochasticity in such systems.

Additionally, humans can engage in mutual adaptation, working collaboratively to resolve misunderstandings or realign expectations. Technical systems, on the other hand, currently lack shared context or the ability to adapt relationally, shift the burden of interpretation entirely onto the user. As a result, any unpredictable behavior in a system appears more alien and untrustworthy. However, should this capability arise, we discuss this as an alternative setup for trustworthy systems in ~\cref{Subsec::Trust in Technological Systems}.

Finally, humans can explain their decisions and actions, even when those actions seem unpredictable. This capacity for reflection and explanation reassures others and helps repair trust when it is threatened. Technical systems often lack the ability to provide clear or accessible rationales for their outputs, leaving users uncertain about the underlying causes of variability. This inability to explain or contextualize behavior amplifies the trust gap. This underscores exactly the decompositional approach that we propose in ~\cref{Sec::Opening the Black Box}.

In summary, the relative stability of human values, the presence of signaling mechanisms, shared social contexts, accountability, and the ability to provide explanations all mitigate the impact of stochasticity in interpersonal relationships. In contrast, technical systems lack these mechanisms, making their stochastic behavior far more likely to erode trust. This fundamental difference highlights the need for intentional design in technical systems to manage stochasticity and preserve user trust.

\end{document}